\documentclass[sigconf]{acmart}
\renewcommand\footnotetextcopyrightpermission[1]{} % removes footnote with conference information in first column
\usepackage{booktabs} % For formal tables
\usepackage{hyperref}
\usepackage{longtable}
\usepackage{enumerate}
\usepackage{graphicx}
\usepackage{subcaption}
\usepackage{xparse}
\usepackage{xr}
\usepackage{amsmath}
\usepackage{algorithm,algpseudocode}
\usepackage{multicol}
\usepackage{float}

\algnewcommand\algorithmicforeach{\textbf{for each}}
\algdef{S}[FOR]{ForEach}[1]{\algorithmicforeach\ #1\ \algorithmicdo}

\PassOptionsToPackage{hyphens}{url}\usepackage{hyperref}

\usepackage{etoolbox}
\patchcmd{\subsubsection}{\bfseries}{\itshape}{}{}

\newcommand\secref[1]{Sec.~\ref{#1}}

\newcommand\appref[1]{App.~\ref{#1}}

\newcommand{\nospace}[1]{\todocolor{[will be removed for space: #1]}}

\patchcmd{\subsubsection}{-.5em}{.5em}{}{}

\setcopyright{none} % No copyright notice required for submissions

\begin{document}
\sloppy
\title[A User-Centric Approach to Quantifying the Privacy of Websites]{By the User, for the User: A User-Centric Approach to Quantifying the Privacy of Websites}
%\title{For the user, by the user: A user-centric approach to quantifying the privacy of websites}

\author{Matius Chairani}
\authornote{Authors contributed equally to this research.}
\affiliation{%
  \institution{Jacobs University Bremen}}
\email{mchairani@jacobs-alumni.de}

\author{Mathieu Chevalley}
\affiliation{%
  \institution{Ecole Polytechnique F\'ed\'erale de Lausanne}}
  \authornotemark[1]
\email{mathieu.chevalley@alumni.epfl.ch}

\author{Abderrahmane Lazraq}
\affiliation{%
  \institution{Ecole Polytechnique F\'ed\'erale de Lausanne}}
  \authornotemark[1]
\email{abderrahmane.lazraq@alumni.epfl.ch}

\author{Sruti Bhagavatula}
\affiliation{%
  \institution{Carnegie Mellon University}}
\email{srutib@cmu.edu}

\begin{abstract}
% \sruti{Third party trackers are commonplace insert number. Lots of previous work has been conducted to detect tracking
% and measuring cookies, etc. This is the first study that quantifies the amount of privacy on websites from a user-centric standpoint.}

Third-party tracking is common on almost all commercially operated websites.
Prior work has studied in detail the extent of third-party tracking on the web,
detection of third-party trackers, and defending against third-party tracking.
Existing research and tools have also attempted to inform web users of trackers
and the extent of their privacy violations. However, existing tools do not
take into account users' perceptions of and understanding of the extent of trackers on the web. Taking these factors into account is important for the usability of such tools so that users can be aware and protect themselves to a reasonable and necessary extent
that aligns with their overall comfort with trackers.

In this paper, we elicit user perceptions and preferences about different trackers on various websites through an online survey of 43 users. We use this data to bootstrap a privacy scoring system. This scoring system weights the usage of trackers and the dispersion of user data within a page to third parties, with the type of website being visited. Our work presents a proof-of-concept methodology and tool to calculate a user-centric privacy score with preliminary bootstrap user data. We conclude with concrete future directions.

% 	- Talk about privacy trackers on websites.
% 	- Lots of prior work has tried to detect trackers on websites
% Ours is the first work to quantify the privacy leakage on websites based on user-elicited data

\end{abstract}
\maketitle

\newcommand{\todocolor}[1]{\unskip}
\newcommand{\sruti}[1]{\todocolor{[[Sruti: #1]]}}

\section{Introduction}
\label{sec:intro}

Third-party trackers (TPT) are present on almost every website~\cite{englehardt2016online}.
TPTs allow websites to track their visitors across multiple digital services
in order to serve more targeted content~\cite{mayer2012third, roesner2012detecting}. While useful, these trackers
induce privacy risks into the websites for its visitors.

Improving the detection of TPTs has been a topic of much
investigation in the research community~\cite{jackson2006protecting, roesner2012detecting}. 
Additionally, browser extensions--e.g., Ghostery, AdBlockPlus~\cite{signanini2014ghostery, adblock}--detect
tracking URLs or ads within a website and display them to the user, sometimes blocking
the URLs from even being loaded. Previous work and existing tools
have been useful to prevent advanced trackers from operating on websites
and to inform users of how much tracking is occurring during their browsing; however, none of these tools are able
to 1) quantify the privacy of a website according to what average users perceive to be sensitive or 2) inform
these users of privacy-violating website components such that they can reasonably assess
a website's privacy in comparison to other websites they visit.

In this paper, we take a step towards being able to
quantify the privacy of a website in a user-centric way. \nospace{the form of a score, based
on user-elicited perceptions of privacy violations on different types of websites.}
We first conduct a survey of 43 participants on Amazon Mechanical Turk~\cite{turk} asking
participants about their awareness of different kinds of trackers, their comfort with these trackers,
and for what purposes they would not be okay with a specific tracker being used by a website.
Using the results of the survey, we
implement a privacy scoring algorithm in a browser extension
that determines which trackers are present on the page running the extension. The extension assigns higher or lower scores
for different tracking components based on the comfort or allowance participants in the study
overall reported with certain types of websites along with the amount of third parties operating the trackers on the page. \sruti{not happy with this sentence}

Our contributions, though preliminary,
provide a proof-of-concept system to quantify privacy in a way meaningful to users, and to empower
users to assess their privacy on websites.

We next provide some background about TPTs (\secref{sec:background})
and related work (\secref{sec:rel-work}). The following two sections describe both 
the user study and results (\secref{sec:user-study}) and the design and implementation of the
privacy scoring mechanism (\secref{sec:priv-score}). We conclude with limitations and future work (\secref{sec:lim-future}).
\section{Background on TPTs}
\label{sec:background}

Third-party trackers (TPTs) are
web components placed onto a webpage and tracked by an external domain.
For example, a known website may include scripts from other domains who
want to track users' behavior on that website, e.g., a TPT in the form of a cookie~\cite{tracking-works}.

\nospace{TPTs are run on a server
completely different  from the server hosting the webpage the tracker is running on.
Therefore, a TPT
has to be separately notified when a user loads a page on the 
website who subscribed to their service. This
is generally done by having the user's browser send an HTTP 
request to the TPT's server every time it loads a page on the tracked site. }
\nospace{TPTs in practice are implemented by incorporating a script written 
by the third-party tracking service. When the webpage is loaded, this 
script gets executed, e.g., it may ask the browser to load a
1x1 pixel image (referred to as a tracking pixel) from the TPT's server. 
The script also can create cookies containing a generated 
session ID~\cite{tracking-works}. In this way, the TPT is able to use the ID to group 
together all requests and web page data that come from the same user.}
%in much the same way as the website requesting the service itself, 
%because the same ID is sent to the TPT every time your browser requests 
%the tracking pixel in the future.

In our work, we study the following eight categories of TPTs according to their purpose and functionality~\cite{signanini2014ghostery}:

\begin{enumerate}
	\item \textbf{Session Replay:} TPTs that track users' view (browser or screen output), user
	input (keyboard and mouse inputs), and logs of network events or console logs.
	
	\item \textbf{Adult advertising:} TPTs that track users' browsing behavior on adult websites for
	adult ad retargeting and behavioral advertising.

	\item \textbf{Social media:} TPTs that study users' browsing behavior on other websites to better
	target users on their social media platform.
	
	\item \textbf{Analytics:} Most general-purpose TPTs that track website usage across different webpages.
	For example, these trackers may examine browsing behavior to empower cross-platform tracking or recording demographics information (e.g. age, gender, location).
	
	\item \textbf{Advertising:} TPTs that track users' browsing behavior on websites for ad retargeting and behavioral advertising.

	\item \textbf{Comments:} TPTs that identify users in comment sections of webpages, including articles and product reviews.
	
	\item \textbf{Audio and video player:} TPTs that track users' behavior when interacting
	with video and audio content.
	
	\item \textbf{Customer interaction:} TPTs that enable e-commerce shops to assist users through a
	pop-up chat box when their mouse is idle while shopping online.

% 	\item \textbf{Website functionality:} TPTs that enable website functions such as privacy notices,
% 	fonts, or tag management.
	
\end{enumerate}
\section{Related Work}
\label{sec:rel-work}

% Third party trackers have been studied on the web extensively in recent years ...
% \par
% Few studies also propose, besides detection of these tracking capabilities, a way to measure the extent of this tracking on different websites during browsing.
% \par
% But to the best of our knowledge, no other study produced a user perception driven scoring mechanism, which is what we attempt to do here. In fact, a scoring mechanism which takes users attitude towards third party trackers as an input is more suitable and more informative to this end user. 

There has been a multitude of prior research studying third-party trackers and defenses against tracking. 

Many researchers have conducted longitudinal studies to study the extent of tracking on the web and the extent of overall privacy-violating web components~\cite{englehardt2016online, roesner2012detecting, krishnamurthy2011privacy, krishnamurthy2009leakage, krishnamurthy2009privacy, krishnamurthy2006generating, englehardt2015cookies, guha2010challenges, ihm2011towards, jackson2006protecting, jensen2007tracking, jang2010empirical, lerner2016internet}. Researchers have also developed a number of methodologies and tools for measuring tracking at this scale~\cite{englehardt2015openwpm}. Beyond this, improving the detection of trackers has been an extensively studied topic~\cite{roesner2012detecting}, also in the context of advertisements~\cite{Orr:2012:AIJ:2381966.2381968, bhagavatula2014leveraging} and privacy-violating web components in general~\cite{krishnamurthy2011privacy, krishnamurthy2009leakage, krishnamurthy2009privacy, krishnamurthy2006generating}.

There is also a large body of work aimed at implementing defenses against web tracking~\cite{roesner2012detecting}. These methods range from privacy-preserving advertisements~\cite{toubiana2010adnostic, fredrikson2010repriv, guha2011privad} to privacy protection on social media~\cite{dhawan2012priv3, kontaxis2012privacy, roesner2012sharemenot}.

With respect to the end-user, there have been a number of studies to understand users' perceptions of web tracking~\cite{leon2013matters, mcdonald2010americans, ur2012smart, tan2018comparing, chanchary2015user, tracking:pets2016}. Additionally, researchers have aimed to comprehensively quantify the privacy of websites and have built tools to usably inform users of violations~\cite{hamed2013evaluation}. These works serve as motivation for our work to design a privacy-quantifying system that takes the users' perceptions into account, which has not been studied before.

% There has been a multitude of prior research studying third party trackers and defenses against such tracking. Researchers have conducted measurements to
% understand the extent of third party tracking on the web and have
% developed tools and methodologies to conduct such longitudinal analyses~\cite{englehardt2015openwpm, englehardt2015cookies, guha2010challenges, ihm2011towards, jackson2006protecting, jensen2007tracking, jang2010empirical}. 
% Researchers have also extensively studied defenses against 
% web tracking~\cite{}. 

% Researchers have also focused on measuring the privacy leaks introduced on the web~\cite{krishnamurthy2011privacy, krishnamurthy2009leakage, krishnamurthy2009privacy, krishnamurthy2006generating}.

% Researchers have studied users' perceptions of tracking on the web~\cite{leon2013matters, mcdonald2010americans, ur2012smart, tan2018comparing}.
\section{Eliciting user perceptions}
\label{sec:user-study}

In this section, we describe our user study to elicit user perceptions of
trackers on different types of websites. 

\subsection{Methodology}

To collect data about what types of tracking users are comfortable with on
different websites, we conducted a survey on Amazon Mechanical Turk~\cite{turk} (approved by our institution's ethics board).
The survey roughly followed this pattern: for each TPT category described in \secref{sec:background} \nospace{except ``Website functionality'' (since we treated TPTs in that category as necessary and therefore, meaningless to ask about comfort with)},
we constructed a scenario describing how companies use the TPTs in the category, without requiring any
prior knowledge from the respondent. For each scenario, we asked participants 1) if they were aware
of the type of tracking in the scenario, 2) about their comfort with knowing that the website runs this kind of 
tracking, and 3) which kind of websites they think should not use the tracking
services described in the scenario (website categories described in Table~\ref{tab:site-cats}).
See \appref{sec:survey} for the full survey.

The survey took less than ten minutes to complete and participants were compensated with \$2.

\subsection{Results}
Our study had 43 participants. 56\% of participants identified as male, the rest identifying as female.
The ages of participants ranged from the 18-30 bracket to the 60+ bracket.

Due to the small sample size, we do not perform any statistical analyses on the responses to
the survey. Furthermore, the primary purpose of the user data is to bootstrap the privacy scoring system (described in
\secref{sec:priv-score}). However,
we report some descriptive statistics of the survey results.

% \begin{table}[]
% \begin{tabular}{|l|l|}
% \hline
% \textit{\textbf{TPT type}} & \textit{\textbf{\% of participants unaware}} \\ \hline
% Session replay & 61\% \\ \hline
% Comment & 58\% \\ \hline
% Analytics & 46\% \\ \hline
% Adult advertising & 42\% \\ \hline
% Audio and video player & 33\% \\ \hline
% Social media & 26\% \\ \hline
% Customer interaction & 16\% \\ \hline
% Advertising & 13\% \\ \hline
% % Website functionality & \sruti{please insert} \\ \hline
% \end{tabular}
% \caption{Percentage of users who were unaware of the type of tracking \label{tab:unaware}}
% \end{table}

Looking at the number of participants who reported to be either ``somewhat uncomfortable'' or 
``very uncomfortable''
with each type of tracker, we rank users' sensitivity to each category and use this ranking of sensitivity to
assign a score to each category used to compute a privacy score (described later in \secref{sec:priv-score}). 
As we hypothesized, session replay is the type of
TPT people are the least comfortable with and customer interaction was the TPT category users were
most comfortable with. 
% we rank the sensitivity of each tracker in Table~\ref{fig:ranking}, i.e., the higher-ranked the tracker,
% the more number of participants answered either of two ``uncomfortable'' options. 
% Based on the results of this survey, after ranking the TPT categories from highest amount of discomfort to lowest,
% we assign a score to each TPT category (whose use we describe in more detail in \secref{sec:priv-score}), where a higher score
% indicates a TPT category with which users indicated higher discomfort (see Table~\ref{I-tab:base-score}). We also assign the lowest score of 1 to ``Website functionality'' TPTs.

Table~\ref{tab:unaware} shows the extent of unawareness for each TPT category. Again, as hypothesized, session replay was also
the least known TPT.

\begin{table}[!htbp]
\begin{tabular}{|l|l|}
\hline
\textit{\textbf{TPT type}} & \textit{\textbf{\% of participants unaware}} \\ \hline
Session replay & 61\% \\ \hline
Comment & 58\% \\ \hline
Analytics & 46\% \\ \hline
Adult advertising & 42\% \\ \hline
Audio and video player & 33\% \\ \hline
Social media & 26\% \\ \hline
Customer interaction & 16\% \\ \hline
Advertising & 13\% \\ \hline
% Website functionality & \sruti{please insert} \\ \hline
\end{tabular}
\caption{Percentage of users who were unaware of the type of tracking \label{tab:unaware}}
\vspace{-2em}
\end{table}

% \begin{table}[]
% \begin{tabular}{|l|l|}
% \hline
% \textit{\textbf{Rank}} & \textit{\textbf{TPT type}} \\ \hline
% 1 & Session replay \\ \hline
% 2 & Adult advertising \\ \hline
% 3 & Social media \\ \hline
% 4 & Analytics \\ \hline
% 5 & Advertising \\ \hline
% 6 & Comment \\ \hline
% 7 & Audio and video player \\ \hline
% 8 & Customer interaction \\ \hline
% \end{tabular}
% \label{fig:ranking}
% \end{table}

\begin{table}[!htbp]
\centering
\begin{tabular}{|l|l|}
\hline
\textbf{Name} & Adobe Audience Manager \\ \hline
\textbf{Pattern}  & demdex.net \\ \hline
\textbf{Category}  &  Advertisement \\ \hline
\textbf{Parent company} & Adobe \\ \hline
\end{tabular}
\caption{Example of stored information about a known TPT \label{tab:known-tpts}}
\vspace{-1em}
\end{table}

\sruti{add text about categories of website concern}
\vspace{-3em}
\section{Building a privacy scoring system}
\label{sec:priv-score}

In this section we describe a privacy scoring mechanism based on the results of the user
study described in Section~\ref{sec:user-study}, wherein a higher privacy score implies a higher amount of privacy. The overall system has two components: a database server and a browser extension which computes the score.

\subsection{Database server}
The database server is built from user-elicited data and general information about websites and TPTs.
The following describes the sets of data stored on the server:
\begin{itemize}
	\item \textbf{Categories of websites:} The server returns one of 11 domain categories (Table~\ref{tab:site-cats}) in response to a request containing the domain of the URL visited by a browser. 
	\item \textbf{URL patterns for known TPT URLs:} The server returns a list of patterns of known TPTs that URLs can be matched against in addition to each TPT's category.
	\item \textbf{Blacklist of TPT types:} The server returns whether a TPT type detected was ``blacklisted'' on the domain category. We determine this blacklist based on the survey results when users indicated that they thought a certain type of
	TPT should not be used on a certain type of domain. For example, users indicated that advertising TPTs should not be used on adult domains, implying  ``advertising'' TPTs are blacklisted for adult domains.
	\nospace{\item \textbf{Existing TPT-based scores:} The server stores all \emph{site-aggregate TPT scores} (described in \secref{sec:extension}) per domain as the browser extension
	computes them.}
\end{itemize}

\begin{table}[!htbp]
\centering
\begin{tabular}{|c|c|}
\hline \textbf{Category} & \textbf{Example} \\ \hline
Adult & www.pornhub.com \\\hline
Banking & www.bankofamerica.com \\\hline
E-commerce & www.underarmour.com \\\hline
Educational & www.cmu.edu \\\hline
Healthcare & www.upmc.com \\\hline
News & www.cnn.com \\\hline
Non-Governmental Organization & www.worldwildlife.org \\\hline
Political & www.donaldjtrump.com \\\hline
Social Media & www.facebook.com \\\hline
Subscription-Based Service & www.spotify.com \\\hline
Other (non e-commerce) & www.sunoco.com \\\hline
\end{tabular}
\caption{Domain categories and examples \label{tab:site-cats}}
\vspace{-2em}
\end{table}

\begin{figure}[!htbp]
\centering
\includegraphics[width=80mm, height=110mm]{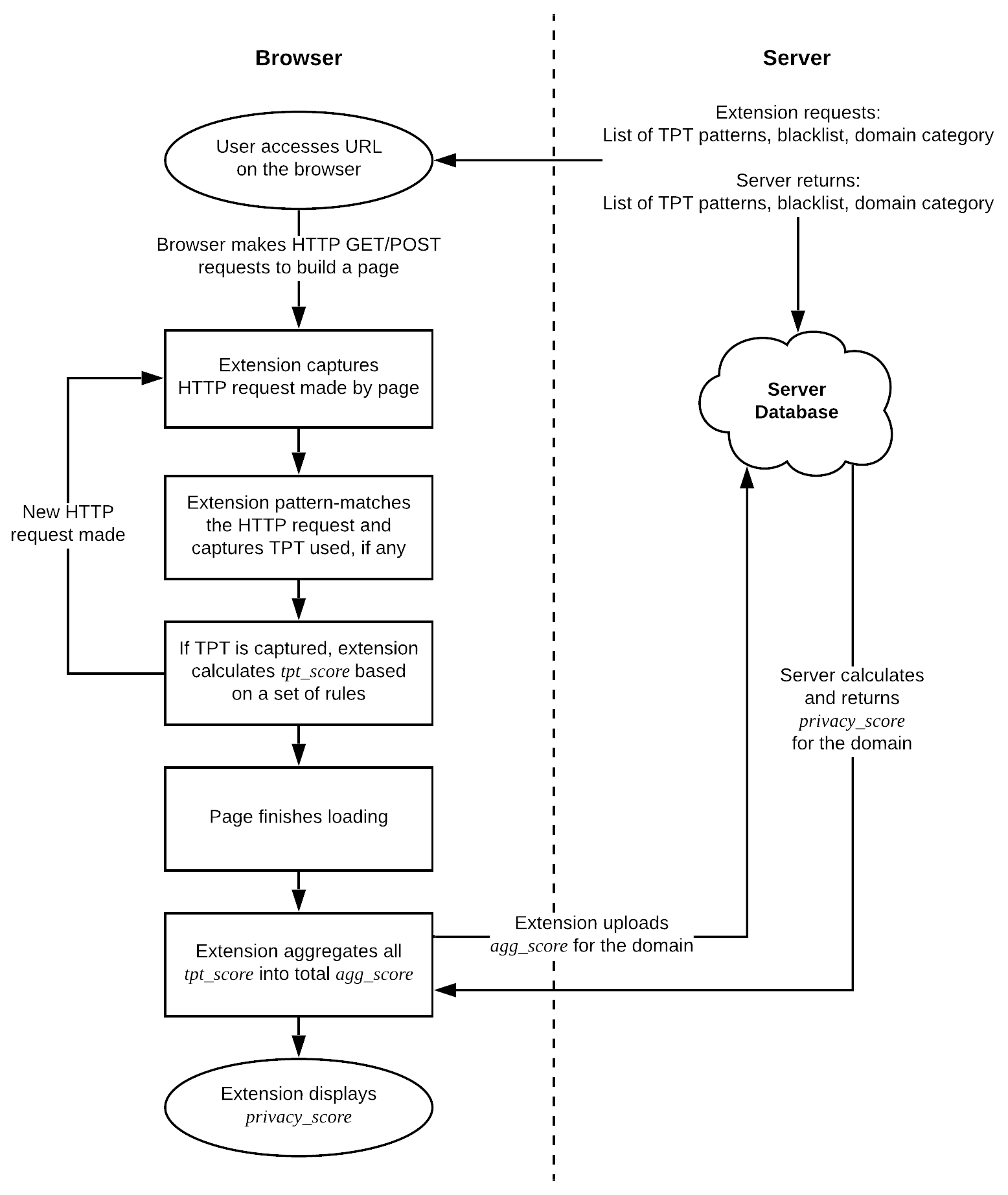}
\captionof{figure}{Flow of the browser extension \label{fig:extension_flow}}
\end{figure}

% \begin{table}[]
% \begin{tabular}{|l|l|}
% \hline
% \textit{\textbf{TPT type}} & \textit{\textbf{Blacklisted for categories}} \\ \hline
% Session Replay & all \\ \hline
% Adult Advertising & - \\ \hline
% Social Media & banking, healthcare \\ \hline
% Analytics & banking \\ \hline
% Advertising & adult, banking, healthcare \\ \hline
% Comments & banking, healthcare, political \\ \hline
% Audio and video player & - \\ \hline
% Customer interaction & adult \\ \hline
% Website functionality & - \\ \hline
% \end{tabular}
% \end{table}

To populate this category mapping, we analyzed 49 safe-for-work websites from the Alexa top 50 sites list in November 2018~\cite{alexa} and manually identified TPT URLs included within the page source. To classify domains, multiple authors browsed the internet normally for three days (recording visited URLs)
and were able to manually label 122 distinct domains according to one of the 11 categories. We further analyzed the source of these 122 domain webpages
to expand our list of TPT matches and ended up with 187 TPT URL patterns\footnote{We construct a URL pattern to be a regex matching a domain and optionally some part of the TPT URL's path. See Table~\ref{tab:known-tpts} for examples of TPT patterns.} in total.

\subsection{Browser extension to compute the privacy score}
\label{sec:extension}
The overall privacy score computed by the extension is based on users' comfort with the usage of different trackers in different contexts in addition to the extent of dispersion of user data within a page to third parties.

First, the browser fetches the category of the domain from the server.
Next, the browser requests a list of known TPT patterns and a list of blacklisted TPTs for the domain category.
After fetching this information, the extension then identifies 
all HTTP requests made by the page that match any of the TPT URL patterns.
A score is computed for each detected TPT. This TPT score is interpreted in the opposite direction of how the overall privacy score should be interpreted, i.e., the TPT score will be higher when the page contains a higher number of privacy-violating components or when users' data is being shared with a higher number of unique third parties, i.e., less privacy. The TPT score is calculated for each TPT URL by first making use of a base score assigned
to the TPT's category. Table~\ref{tab:base-score} contains the listing of base scores for each TPT category based
on the ranking of comfort described in \secref{sec:user-study}\nospace {where the lowest score is also assigned to ``Website functionality''}. The TPT score is first initialized to the base score for the TPT category. If the TPT category is blacklisted for the domain category, the TPT score is increased by a factor of 1.5. The extension then checks if the company operating the TPT in question operating more than one TPT on the same web page. If so, we interpret the user's data to have a lower extent of dispersion (since this limits the amount of unique domains tracking user data) and decrease the TPT score
by 1. The computed TPT scores for each TPT are then added together to form an aggregate TPT score for the page ($agg\_score$).

The extension sends $agg\_score$ and the domain's category to the server. The
server compares this score against 1) previously computed scores for all the other domains in the category of the domain in question and 2) previously computed scores for all domains seen before. We compute the following two percentile values based on the scores in the above two groups: \emph{categorical percentile}:, i.e., the percentile of the browser-calculated score with respect to how many scores in the first category the calculated score is higher than (i.e., exhibits less privacy than); and a \emph{global percentile}, i.e., similar to the categorical percentile except compared with the scores in the second category. The two percentile values are averaged and the final privacy score is computed by subtracting this average from 100, giving us a privacy score percentage between 1 and 100 (see Algorithm~\ref{alg:score} for a full description).

The flow of the browser extension is shown in Figure~\ref{fig:extension_flow}. \appref{app:sshots} contains screenshots of the browser extension
in action, which we named ``Cookie Police''. In addition to a score, the extension reports how the website compares to other websites in its category and other sites, as well as the companies operating trackers on the site.

\begin{table}[!htbp]
\centering
\begin{tabular}{|c|c|}
\hline \textbf{TPT Type} & \textbf{Base score} \\
\hline
Session Replay & 8 \\
\hline
Adult Advertising & 7 \\
\hline
Social Media & 6 \\
\hline
Analytics & 5 \\
\hline
Advertising & 4 \\
\hline
Comments & 3 \\
\hline
Audio and Video Player & 2 \\
\hline
Customer Interaction & 1 \\
\hline
%Website Functionality & 1 \\ \hline
\end{tabular}
\caption{Base score for each TPT category \label{tab:base-score}}
\vspace{-2em}
\end{table}

\begin{algorithm*}
    \begin{algorithmic}[1]
    \Function{calcTPTScore}{$TPT$, $blacklist$, $seen\_companies$}
    \State $tpt\_score = getBaseCategoryScore(TPT[category])$
    \If{$TPT[category] \in blacklist$}
        \State $tpt\_score = tpt\_score \times 1.5$
    \EndIf
    \If {$TPT[company] \in seen\_companies$}
        \State $tpt\_score = tpt\_score - 1$
    \Else
        \State Add $TPT[company]$ to $seen\_companies$ \algorithmiccomment{Object reference updated}
    \EndIf
    \State \Return $tpt\_score$
    \EndFunction \\ \\
    
    % \Function{computeCP}{category, agg\_score}
    %     \State percentile = 0
    %     \ForAll{score \in domain\_cat\_scores}
    %         \If agg\_score > score}
    %     \EndFor
    %     % % \If {agg\_score > score}
    %     % %         %\State percentile = percentile + 1
    %     % % \EndIf
        
    % %     \State percentile = percentile / |domain\_cat\_scores|
    % % \State \Return percentile
    % \EndFunction 
    
    \Comment{\textbf{-----------------------------------------------------------------------Execution starts here}}
    \State $domain\_cat$, $blacklist$, $TPT\_patterns$ = $fetchInfoFromServer(domain)$
    \State $seen\_companies = \{\}$
    \State $agg\_score = 0$
    
    \ForEach{$TPT \in page\_load$} \algorithmiccomment{TPT object contains: parent company and TPT category}
        \State $tpt\_score = calcTPTScore(TPT, blacklist, seen\_companies)$
        \State $agg\_score = agg\_score + tpt\_score$
    \EndFor \\

    \State $cat\_percentile = computeCategoricalPercentile(TPT[category], agg\_score)$ \algorithmiccomment{\% of previously seen domains (across users) in the category that agg\_score is higher than} \\
    \State $glob\_percentile = computeGlobalPercentile(TPT[category], agg\_score)$
    \algorithmiccomment{\% of all previously seen domains (across users) that agg\_score is higher than} \\

    \State $privacy\_score = 100 - (cat\_percentile + glob\_percentile) / 2$
    
    \caption{Computing the privacy score \label{alg:score}}
    \end{algorithmic}
\end{algorithm*}
\section{Limitations and Future work}
\label{sec:lim-future}

Our work is subject to limitations due to limited resources and budget and therefore,
lends to ample future work. \nospace{We describe these below in 
addition to how the limitations pave way to future work as well as other future 
directions.}

The sample size of survey respondents was not large enough to build a representative 
model of user perceptions used to compute privacy scores or to study small effects of 
statistical significance. Additionally, our respondents were all US residents. Future work should study perceptions of tracking privacy on a
large scale, possibly even across countries, and build the scoring system based on more representative data. Alternatively, the score can be computed on an individual basis or on a population basis (e.g., per country), i.e., the score for one user or population is uniquely based on data collected by that user or population.

We assigned a category to a domain by manual labeling of a small set of domains. If the scoring extension is deployed in the wild, this manual categorization is not scalable. Future work should
use an automatic categorization such as Amazon's Alexa Web Information Services (AWIS)~\cite{awis}, SimilarWeb's Website Categorization API~\cite{sim-web-classification}, or Google's natural language content classification~\cite{g-content} and accordingly update the
user study questions with the categories of interest. Furthermore, there is no 
existing database of recognized TPT URL patterns. Therefore, we use the 
labels produced by other tracking detection tools on a set of URLs as our ground truth for generating URL patterns. Future work could study how to automatically generate TPT-matching regular expressions from historical data and other tools, which automatically update over time. Additionally, the research community could benefit from collectively maintaining a public database of known TPT URL patterns implemented by various extensions.

The scoring algorithm is based on aggregate user perception. This lends 
to possible deviations from the perceptions of a specific user who might 
have outlying perceptions. Future work could implement such a scoring 
extension with a feedback mechanism, wherein users are able to provide 
feedback on certain aspects of the score or provide custom answers to some 
of the questions in the user study.

Future work should involve usability testing and a systematic 
evaluation of accuracy of a more sophisticated version of our 
tool. This would involve testing general usability, testing 
a feedback mechanism as mentioned above, evaluating the accuracy 
of the TPT filters and evaluating the extent of evasion of the TPT patterns in the wild. Related to this, future work should ensure that the detection of TPT URLs is comparable to the detection by state-of-the-art tools (e.g. Ghostery) or should use the detection system employed by such tools either through any APIs or open-source implementations.


 % TODO: replace with your keywords

\bibliographystyle{ACM-Reference-Format}
\bibliography{references}


\begin{thebibliography}{38}

%%% the Web address.

\ifx \showCODEN    \undefined \def \showCODEN     #1{\unskip}     \fi
\ifx \showDOI      \undefined \def \showDOI       #1{#1}\fi
\ifx \showISBNx    \undefined \def \showISBNx     #1{\unskip}     \fi
\ifx \showISBNxiii \undefined \def \showISBNxiii  #1{\unskip}     \fi
\ifx \showISSN     \undefined \def \showISSN      #1{\unskip}     \fi
\ifx \showLCCN     \undefined \def \showLCCN      #1{\unskip}     \fi
\ifx \shownote     \undefined \def \shownote      #1{#1}          \fi
\ifx \showarticletitle \undefined \def \showarticletitle #1{#1}   \fi
\ifx \showURL      \undefined \def \showURL       {\relax}        \fi
% invisible to TeX
\providecommand\bibfield[2]{#2}
\providecommand\bibinfo[2]{#2}
\providecommand\natexlab[1]{#1}
\providecommand\showeprint[2][]{arXiv:#2}

\bibitem[\protect\citeauthoryear{??}{tra}{2017}]%
        {tracking-works}
 \bibinfo{year}{2017}\natexlab{}.
\newblock \showarticletitle{How does online tracking actually work?}
\newblock \bibinfo{journal}{\emph{Robert Heaton}} (\bibinfo{year}{2017}).
\newblock
\urldef\tempurl%
\url{https://robertheaton.com/2017/11/20/how-does-online-tracking-actually-work/}
\showURL{%
\tempurl}


\bibitem[\protect\citeauthoryear{??}{ale}{2018}]%
        {alexa}
 \bibinfo{year}{2018}\natexlab{}.
\newblock \bibinfo{title}{The top 500 sites on the web The sites in the top
  sites lists are ordered by their 1 month Alexa traffic rank.The 1 month rank
  is calculated using a combination of average daily visitors and pageviews
  over the past month. The site with the highest combination of visitors and
  pageviews is ranked \#1}.
\newblock
\newblock
\urldef\tempurl%
\url{https://www.alexa.com/topsites}
\showURL{%
\tempurl}


\bibitem[\protect\citeauthoryear{??}{adb}{2019}]%
        {adblock}
 \bibinfo{year}{2019}\natexlab{}.
\newblock \showarticletitle{Adblock Plus: The world's No. 1 free ad blocker}.
\newblock \bibinfo{journal}{\emph{Adblock Plus | The world's \# 1 free ad
  blocker}} (\bibinfo{year}{2019}).
\newblock
\urldef\tempurl%
\url{https://adblockplus.org/en/}
\showURL{%
\tempurl}


\bibitem[\protect\citeauthoryear{??}{tur}{2019}]%
        {turk}
 \bibinfo{year}{2019}\natexlab{}.
\newblock \showarticletitle{Amazon Mechanical Turk}.
\newblock \bibinfo{journal}{\emph{Amazon Mechanical Turk}}
  (\bibinfo{year}{2019}).
\newblock
\urldef\tempurl%
\url{https://www.mturk.com/}
\showURL{%
\tempurl}


\bibitem[\protect\citeauthoryear{??}{awi}{2019}]%
        {awis}
 \bibinfo{year}{2019}\natexlab{}.
\newblock \showarticletitle{AWIS | Alexa Web Information Service - Traffic
  Metrics for any Website}.
\newblock  (\bibinfo{year}{2019}).
\newblock
\urldef\tempurl%
\url{https://aws.amazon.com/awis/}
\showURL{%
\tempurl}


\bibitem[\protect\citeauthoryear{??}{g-c}{2019}]%
        {g-content}
 \bibinfo{year}{2019}\natexlab{}.
\newblock \showarticletitle{Google | Cloud Natural Language API: Content
  Categories}.
\newblock  (\bibinfo{year}{2019}).
\newblock
\urldef\tempurl%
\url{https://cloud.google.com/natural-language/docs/categories}
\showURL{%
\tempurl}


\bibitem[\protect\citeauthoryear{??}{sim}{2019}]%
        {sim-web-classification}
 \bibinfo{year}{2019}\natexlab{}.
\newblock \showarticletitle{SimilarWeb | Website Categorization API}.
\newblock  (\bibinfo{year}{2019}).
\newblock
\urldef\tempurl%
\url{https://www.similarweb.com/corp/developer/website_categorization_API}
\showURL{%
\tempurl}


\bibitem[\protect\citeauthoryear{Bhagavatula, Dunn, Kanich, Gupta, and
  Ziebart}{Bhagavatula et~al\mbox{.}}{2014}]%
        {bhagavatula2014leveraging}
\bibfield{author}{\bibinfo{person}{Sruti Bhagavatula},
  \bibinfo{person}{Christopher Dunn}, \bibinfo{person}{Chris Kanich},
  \bibinfo{person}{Minaxi Gupta}, {and} \bibinfo{person}{Brian Ziebart}.}
  \bibinfo{year}{2014}\natexlab{}.
\newblock \showarticletitle{Leveraging machine learning to improve unwanted
  resource filtering}. In \bibinfo{booktitle}{\emph{Proceedings of the 2014
  Workshop on Artificial Intelligent and Security Workshop}}. ACM,
  \bibinfo{pages}{95--102}.
\newblock


\bibitem[\protect\citeauthoryear{Chanchary and Chiasson}{Chanchary and
  Chiasson}{2015}]%
        {chanchary2015user}
\bibfield{author}{\bibinfo{person}{Farah Chanchary} {and}
  \bibinfo{person}{Sonia Chiasson}.} \bibinfo{year}{2015}\natexlab{}.
\newblock \showarticletitle{User perceptions of sharing, advertising, and
  tracking}. In \bibinfo{booktitle}{\emph{Eleventh Symposium On Usable Privacy
  and Security ($\{$SOUPS$\}$ 2015)}}. \bibinfo{pages}{53--67}.
\newblock


\bibitem[\protect\citeauthoryear{Dhawan, Kreibich, and Weaver}{Dhawan
  et~al\mbox{.}}{2012}]%
        {dhawan2012priv3}
\bibfield{author}{\bibinfo{person}{Mohan Dhawan}, \bibinfo{person}{Christian
  Kreibich}, {and} \bibinfo{person}{Nicholas Weaver}.}
  \bibinfo{year}{2012}\natexlab{}.
\newblock \showarticletitle{Priv3: A third party cookie policy}. In
  \bibinfo{booktitle}{\emph{W3C Workshop: Do Not Track and Beyond}}.
\newblock


\bibitem[\protect\citeauthoryear{Englehardt, Eubank, Zimmerman, Reisman, and
  Narayanan}{Englehardt et~al\mbox{.}}{2015a}]%
        {englehardt2015openwpm}
\bibfield{author}{\bibinfo{person}{Steven Englehardt}, \bibinfo{person}{Chris
  Eubank}, \bibinfo{person}{Peter Zimmerman}, \bibinfo{person}{Dillon Reisman},
  {and} \bibinfo{person}{Arvind Narayanan}.} \bibinfo{year}{2015}\natexlab{a}.
\newblock \showarticletitle{OpenWPM: An automated platform for web privacy
  measurement}.
\newblock \bibinfo{journal}{\emph{Manuscript, mar}} (\bibinfo{year}{2015}).
\newblock


\bibitem[\protect\citeauthoryear{Englehardt and Narayanan}{Englehardt and
  Narayanan}{2016}]%
        {englehardt2016online}
\bibfield{author}{\bibinfo{person}{Steven Englehardt} {and}
  \bibinfo{person}{Arvind Narayanan}.} \bibinfo{year}{2016}\natexlab{}.
\newblock \showarticletitle{Online tracking: A 1-million-site measurement and
  analysis}. In \bibinfo{booktitle}{\emph{Proceedings of the 2016 ACM SIGSAC
  conference on computer and communications security}}. ACM,
  \bibinfo{pages}{1388--1401}.
\newblock


\bibitem[\protect\citeauthoryear{Englehardt, Reisman, Eubank, Zimmerman, Mayer,
  Narayanan, and Felten}{Englehardt et~al\mbox{.}}{2015b}]%
        {englehardt2015cookies}
\bibfield{author}{\bibinfo{person}{Steven Englehardt}, \bibinfo{person}{Dillon
  Reisman}, \bibinfo{person}{Christian Eubank}, \bibinfo{person}{Peter
  Zimmerman}, \bibinfo{person}{Jonathan Mayer}, \bibinfo{person}{Arvind
  Narayanan}, {and} \bibinfo{person}{Edward~W Felten}.}
  \bibinfo{year}{2015}\natexlab{b}.
\newblock \showarticletitle{Cookies that give you away: The surveillance
  implications of web tracking}. In \bibinfo{booktitle}{\emph{Proceedings of
  the 24th International Conference on World Wide Web}}. International World
  Wide Web Conferences Steering Committee, \bibinfo{pages}{289--299}.
\newblock


\bibitem[\protect\citeauthoryear{Fredrikson and Livshits}{Fredrikson and
  Livshits}{2010}]%
        {fredrikson2010repriv}
\bibfield{author}{\bibinfo{person}{Matthew Fredrikson} {and}
  \bibinfo{person}{Benjamin Livshits}.} \bibinfo{year}{2010}\natexlab{}.
\newblock \showarticletitle{RePriv: Re-envisioning in-browser privacy}. In
  \bibinfo{booktitle}{\emph{Proc. IEEE Symp. Security, Privacy (SP)(May
  2011)}}.
\newblock


\bibitem[\protect\citeauthoryear{Guha, Cheng, and Francis}{Guha
  et~al\mbox{.}}{2010}]%
        {guha2010challenges}
\bibfield{author}{\bibinfo{person}{Saikat Guha}, \bibinfo{person}{Bin Cheng},
  {and} \bibinfo{person}{Paul Francis}.} \bibinfo{year}{2010}\natexlab{}.
\newblock \showarticletitle{Challenges in measuring online advertising
  systems}. In \bibinfo{booktitle}{\emph{Proceedings of the 10th ACM SIGCOMM
  conference on Internet measurement}}. ACM, \bibinfo{pages}{81--87}.
\newblock


\bibitem[\protect\citeauthoryear{Guha, Cheng, and Francis}{Guha
  et~al\mbox{.}}{2011}]%
        {guha2011privad}
\bibfield{author}{\bibinfo{person}{Saikat Guha}, \bibinfo{person}{Bin Cheng},
  {and} \bibinfo{person}{Paul Francis}.} \bibinfo{year}{2011}\natexlab{}.
\newblock \showarticletitle{Privad: Practical privacy in online advertising}.
  In \bibinfo{booktitle}{\emph{USENIX conference on Networked systems design
  and implementation}}. \bibinfo{pages}{169--182}.
\newblock


\bibitem[\protect\citeauthoryear{Hamed, Ayed, Kaafar, and Kharraz}{Hamed
  et~al\mbox{.}}{2013}]%
        {hamed2013evaluation}
\bibfield{author}{\bibinfo{person}{Asma Hamed}, \bibinfo{person}{Hella
  Kaffel-Ben Ayed}, \bibinfo{person}{Mohamed~Ali Kaafar}, {and}
  \bibinfo{person}{Ahmed Kharraz}.} \bibinfo{year}{2013}\natexlab{}.
\newblock \showarticletitle{Evaluation of third party tracking on the web}. In
  \bibinfo{booktitle}{\emph{8th International Conference for Internet
  Technology and Secured Transactions (ICITST-2013)}}. IEEE,
  \bibinfo{pages}{471--477}.
\newblock


\bibitem[\protect\citeauthoryear{Ihm and Pai}{Ihm and Pai}{2011}]%
        {ihm2011towards}
\bibfield{author}{\bibinfo{person}{Sunghwan Ihm} {and} \bibinfo{person}{Vivek~S
  Pai}.} \bibinfo{year}{2011}\natexlab{}.
\newblock \showarticletitle{Towards understanding modern web traffic}. In
  \bibinfo{booktitle}{\emph{Proceedings of the 2011 ACM SIGCOMM conference on
  Internet measurement conference}}. ACM, \bibinfo{pages}{295--312}.
\newblock


\bibitem[\protect\citeauthoryear{Jackson, Bortz, Boneh, and Mitchell}{Jackson
  et~al\mbox{.}}{2006}]%
        {jackson2006protecting}
\bibfield{author}{\bibinfo{person}{Collin Jackson}, \bibinfo{person}{Andrew
  Bortz}, \bibinfo{person}{Dan Boneh}, {and} \bibinfo{person}{John~C
  Mitchell}.} \bibinfo{year}{2006}\natexlab{}.
\newblock \showarticletitle{Protecting browser state from web privacy attacks}.
  In \bibinfo{booktitle}{\emph{Proceedings of the 15th international conference
  on World Wide Web}}. ACM, \bibinfo{pages}{737--744}.
\newblock


\bibitem[\protect\citeauthoryear{Jang, Jhala, Lerner, and Shacham}{Jang
  et~al\mbox{.}}{2010}]%
        {jang2010empirical}
\bibfield{author}{\bibinfo{person}{Dongseok Jang}, \bibinfo{person}{Ranjit
  Jhala}, \bibinfo{person}{Sorin Lerner}, {and} \bibinfo{person}{Hovav
  Shacham}.} \bibinfo{year}{2010}\natexlab{}.
\newblock \showarticletitle{An empirical study of privacy-violating information
  flows in JavaScript web applications}. In
  \bibinfo{booktitle}{\emph{Proceedings of the 17th ACM conference on Computer
  and communications security}}. ACM, \bibinfo{pages}{270--283}.
\newblock


\bibitem[\protect\citeauthoryear{Jensen, Sarkar, Jensen, and Potts}{Jensen
  et~al\mbox{.}}{2007}]%
        {jensen2007tracking}
\bibfield{author}{\bibinfo{person}{Carlos Jensen}, \bibinfo{person}{Chandan
  Sarkar}, \bibinfo{person}{Christian Jensen}, {and} \bibinfo{person}{Colin
  Potts}.} \bibinfo{year}{2007}\natexlab{}.
\newblock \showarticletitle{Tracking website data-collection and privacy
  practices with the iWatch web crawler}. In
  \bibinfo{booktitle}{\emph{Proceedings of the 3rd symposium on Usable privacy
  and security}}. ACM, \bibinfo{pages}{29--40}.
\newblock


\bibitem[\protect\citeauthoryear{Kontaxis, Polychronakis, Keromytis, and
  Markatos}{Kontaxis et~al\mbox{.}}{2012}]%
        {kontaxis2012privacy}
\bibfield{author}{\bibinfo{person}{Georgios Kontaxis},
  \bibinfo{person}{Michalis Polychronakis}, \bibinfo{person}{Angelos~D
  Keromytis}, {and} \bibinfo{person}{Evangelos~P Markatos}.}
  \bibinfo{year}{2012}\natexlab{}.
\newblock \showarticletitle{Privacy-preserving social plugins}. In
  \bibinfo{booktitle}{\emph{Presented as part of the 21st $\{$USENIX$\}$
  Security Symposium ($\{$USENIX$\}$ Security 12)}}. \bibinfo{pages}{631--646}.
\newblock


\bibitem[\protect\citeauthoryear{Krishnamurthy, Naryshkin, and
  Wills}{Krishnamurthy et~al\mbox{.}}{2011}]%
        {krishnamurthy2011privacy}
\bibfield{author}{\bibinfo{person}{Balachander Krishnamurthy},
  \bibinfo{person}{Konstantin Naryshkin}, {and} \bibinfo{person}{Craig Wills}.}
  \bibinfo{year}{2011}\natexlab{}.
\newblock \showarticletitle{Privacy leakage vs. protection measures: the
  growing disconnect}. In \bibinfo{booktitle}{\emph{Proceedings of the Web}},
  Vol.~\bibinfo{volume}{2}. \bibinfo{pages}{1--10}.
\newblock


\bibitem[\protect\citeauthoryear{Krishnamurthy and Wills}{Krishnamurthy and
  Wills}{2009a}]%
        {krishnamurthy2009privacy}
\bibfield{author}{\bibinfo{person}{Balachander Krishnamurthy} {and}
  \bibinfo{person}{Craig Wills}.} \bibinfo{year}{2009}\natexlab{a}.
\newblock \showarticletitle{Privacy diffusion on the web: a longitudinal
  perspective}. In \bibinfo{booktitle}{\emph{Proceedings of the 18th
  international conference on World wide web}}. ACM, \bibinfo{pages}{541--550}.
\newblock


\bibitem[\protect\citeauthoryear{Krishnamurthy and Wills}{Krishnamurthy and
  Wills}{2006}]%
        {krishnamurthy2006generating}
\bibfield{author}{\bibinfo{person}{Balachander Krishnamurthy} {and}
  \bibinfo{person}{Craig~E Wills}.} \bibinfo{year}{2006}\natexlab{}.
\newblock \showarticletitle{Generating a privacy footprint on the internet}. In
  \bibinfo{booktitle}{\emph{Proceedings of the 6th ACM SIGCOMM conference on
  Internet measurement}}. ACM, \bibinfo{pages}{65--70}.
\newblock


\bibitem[\protect\citeauthoryear{Krishnamurthy and Wills}{Krishnamurthy and
  Wills}{2009b}]%
        {krishnamurthy2009leakage}
\bibfield{author}{\bibinfo{person}{Balachander Krishnamurthy} {and}
  \bibinfo{person}{Craig~E Wills}.} \bibinfo{year}{2009}\natexlab{b}.
\newblock \showarticletitle{On the leakage of personally identifiable
  information via online social networks}. In
  \bibinfo{booktitle}{\emph{Proceedings of the 2nd ACM workshop on Online
  social networks}}. ACM, \bibinfo{pages}{7--12}.
\newblock


\bibitem[\protect\citeauthoryear{Leon, Ur, Wang, Sleeper, Balebako, Shay,
  Bauer, Christodorescu, and Cranor}{Leon et~al\mbox{.}}{2013}]%
        {leon2013matters}
\bibfield{author}{\bibinfo{person}{Pedro~Giovanni Leon}, \bibinfo{person}{Blase
  Ur}, \bibinfo{person}{Yang Wang}, \bibinfo{person}{Manya Sleeper},
  \bibinfo{person}{Rebecca Balebako}, \bibinfo{person}{Richard Shay},
  \bibinfo{person}{Lujo Bauer}, \bibinfo{person}{Mihai Christodorescu}, {and}
  \bibinfo{person}{Lorrie~Faith Cranor}.} \bibinfo{year}{2013}\natexlab{}.
\newblock \showarticletitle{What matters to users?: factors that affect users'
  willingness to share information with online advertisers}. In
  \bibinfo{booktitle}{\emph{Proceedings of the ninth symposium on usable
  privacy and security}}. ACM, \bibinfo{pages}{7}.
\newblock


\bibitem[\protect\citeauthoryear{Lerner, Simpson, Kohno, and Roesner}{Lerner
  et~al\mbox{.}}{2016}]%
        {lerner2016internet}
\bibfield{author}{\bibinfo{person}{Adam Lerner}, \bibinfo{person}{Anna~Kornfeld
  Simpson}, \bibinfo{person}{Tadayoshi Kohno}, {and} \bibinfo{person}{Franziska
  Roesner}.} \bibinfo{year}{2016}\natexlab{}.
\newblock \showarticletitle{Internet jones and the raiders of the lost
  trackers: An archaeological study of web tracking from 1996 to 2016}. In
  \bibinfo{booktitle}{\emph{25th $\{$USENIX$\}$ Security Symposium
  ($\{$USENIX$\}$ Security 16)}}.
\newblock


\bibitem[\protect\citeauthoryear{Mayer and Mitchell}{Mayer and
  Mitchell}{2012}]%
        {mayer2012third}
\bibfield{author}{\bibinfo{person}{Jonathan~R Mayer} {and}
  \bibinfo{person}{John~C Mitchell}.} \bibinfo{year}{2012}\natexlab{}.
\newblock \showarticletitle{Third-party web tracking: Policy and technology}.
  In \bibinfo{booktitle}{\emph{2012 IEEE symposium on security and privacy}}.
  IEEE, \bibinfo{pages}{413--427}.
\newblock


\bibitem[\protect\citeauthoryear{McDonald and Cranor}{McDonald and
  Cranor}{2010}]%
        {mcdonald2010americans}
\bibfield{author}{\bibinfo{person}{Aleecia~M McDonald} {and}
  \bibinfo{person}{Lorrie~Faith Cranor}.} \bibinfo{year}{2010}\natexlab{}.
\newblock \showarticletitle{Americans' attitudes about internet behavioral
  advertising practices}. In \bibinfo{booktitle}{\emph{Proceedings of the 9th
  annual ACM workshop on Privacy in the electronic society}}. ACM,
  \bibinfo{pages}{63--72}.
\newblock


\bibitem[\protect\citeauthoryear{Melicher, Sharif, Tan, Bauer, Christodorescu,
  and Leon}{Melicher et~al\mbox{.}}{2016}]%
        {tracking:pets2016}
\bibfield{author}{\bibinfo{person}{William Melicher}, \bibinfo{person}{Mahmood
  Sharif}, \bibinfo{person}{Joshua Tan}, \bibinfo{person}{Lujo Bauer},
  \bibinfo{person}{Mihai Christodorescu}, {and} \bibinfo{person}{Pedro~Giovanni
  Leon}.} \bibinfo{year}{2016}\natexlab{}.
\newblock \showarticletitle{({Do} not) {Track} me sometimes: {Users'}
  contextual preferences for web tracking}.
\newblock \bibinfo{journal}{\emph{Proceedings on Privacy Enhancing
  Technologies}} (\bibinfo{date}{April} \bibinfo{year}{2016}).
\newblock
Issue 2.
\urldef\tempurl%
\url{https://doi.org/10.1515/popets-2016-0009}
\showDOI{\tempurl}


\bibitem[\protect\citeauthoryear{Orr, Chauhan, Gupta, Frisz, and Dunn}{Orr
  et~al\mbox{.}}{2012}]%
        {Orr:2012:AIJ:2381966.2381968}
\bibfield{author}{\bibinfo{person}{Caitlin~R. Orr}, \bibinfo{person}{Arun
  Chauhan}, \bibinfo{person}{Minaxi Gupta}, \bibinfo{person}{Christopher~J.
  Frisz}, {and} \bibinfo{person}{Christopher~W. Dunn}.}
  \bibinfo{year}{2012}\natexlab{}.
\newblock \showarticletitle{An Approach for Identifying JavaScript-loaded
  Advertisements Through Static Program Analysis}. In
  \bibinfo{booktitle}{\emph{Proceedings of the 2012 ACM Workshop on Privacy in
  the Electronic Society}} \emph{(\bibinfo{series}{WPES '12})}.
  \bibinfo{publisher}{ACM}, \bibinfo{address}{New York, NY, USA},
  \bibinfo{pages}{1--12}.
\newblock
\showISBNx{978-1-4503-1663-7}
\urldef\tempurl%
\url{https://doi.org/10.1145/2381966.2381968}
\showDOI{\tempurl}


\bibitem[\protect\citeauthoryear{Roesner, Kohno, and Wetherall}{Roesner
  et~al\mbox{.}}{2012a}]%
        {roesner2012detecting}
\bibfield{author}{\bibinfo{person}{Franziska Roesner},
  \bibinfo{person}{Tadayoshi Kohno}, {and} \bibinfo{person}{David Wetherall}.}
  \bibinfo{year}{2012}\natexlab{a}.
\newblock \showarticletitle{Detecting and defending against third-party
  tracking on the web}. In \bibinfo{booktitle}{\emph{Proceedings of the 9th
  USENIX conference on Networked Systems Design and Implementation}}. USENIX
  Association, \bibinfo{pages}{12--12}.
\newblock


\bibitem[\protect\citeauthoryear{Roesner, Rovillos, Kohno, and
  Wetherall}{Roesner et~al\mbox{.}}{2012b}]%
        {roesner2012sharemenot}
\bibfield{author}{\bibinfo{person}{Franziska Roesner},
  \bibinfo{person}{Christopher Rovillos}, \bibinfo{person}{Tadayoshi Kohno},
  {and} \bibinfo{person}{David Wetherall}.} \bibinfo{year}{2012}\natexlab{b}.
\newblock \showarticletitle{Sharemenot: Balancing privacy and functionality of
  third-party social widgets}.
\newblock \bibinfo{journal}{\emph{Usenix; login}} (\bibinfo{year}{2012}).
\newblock


\bibitem[\protect\citeauthoryear{Signanini and McDermott}{Signanini and
  McDermott}{2014}]%
        {signanini2014ghostery}
\bibfield{author}{\bibinfo{person}{JM Signanini} {and} \bibinfo{person}{B
  McDermott}.} \bibinfo{year}{2014}\natexlab{}.
\newblock \showarticletitle{Ghostery}.
\newblock  (\bibinfo{year}{2014}).
\newblock


\bibitem[\protect\citeauthoryear{Tan, Sharif, Bhagavatula, Beckerle, Mazurek,
  and Bauer}{Tan et~al\mbox{.}}{2018}]%
        {tan2018comparing}
\bibfield{author}{\bibinfo{person}{Joshua Tan}, \bibinfo{person}{Mahmood
  Sharif}, \bibinfo{person}{Sruti Bhagavatula}, \bibinfo{person}{Matthias
  Beckerle}, \bibinfo{person}{Michelle~L Mazurek}, {and} \bibinfo{person}{Lujo
  Bauer}.} \bibinfo{year}{2018}\natexlab{}.
\newblock \showarticletitle{Comparing Hypothetical and Realistic Privacy
  Valuations}. In \bibinfo{booktitle}{\emph{Proceedings of the 2018 Workshop on
  Privacy in the Electronic Society}}. ACM, \bibinfo{pages}{168--182}.
\newblock


\bibitem[\protect\citeauthoryear{Toubiana, Narayanan, Boneh, Nissenbaum, and
  Barocas}{Toubiana et~al\mbox{.}}{2010}]%
        {toubiana2010adnostic}
\bibfield{author}{\bibinfo{person}{Vincent Toubiana}, \bibinfo{person}{Arvind
  Narayanan}, \bibinfo{person}{Dan Boneh}, \bibinfo{person}{Helen Nissenbaum},
  {and} \bibinfo{person}{Solon Barocas}.} \bibinfo{year}{2010}\natexlab{}.
\newblock \showarticletitle{Adnostic: Privacy preserving targeted advertising}.
  In \bibinfo{booktitle}{\emph{Proceedings Network and Distributed System
  Symposium}}.
\newblock


\bibitem[\protect\citeauthoryear{Ur, Leon, Cranor, Shay, and Wang}{Ur
  et~al\mbox{.}}{2012}]%
        {ur2012smart}
\bibfield{author}{\bibinfo{person}{Blase Ur}, \bibinfo{person}{Pedro~Giovanni
  Leon}, \bibinfo{person}{Lorrie~Faith Cranor}, \bibinfo{person}{Richard Shay},
  {and} \bibinfo{person}{Yang Wang}.} \bibinfo{year}{2012}\natexlab{}.
\newblock \showarticletitle{Smart, useful, scary, creepy: perceptions of online
  behavioral advertising}. In \bibinfo{booktitle}{\emph{proceedings of the
  eighth symposium on usable privacy and security}}. ACM, \bibinfo{pages}{4}.
\newblock


\end{thebibliography}

\appendix
 \section{Survey}
\label{sec:survey}

With the rapid development and growth of the Internet, marketers are turning from physical marketing techniques (e.g. sending flyers to your home) to online marketing techniques. With online marketing techniques, marketers are better able to target their advertising to customers, thus reducing users annoyance from receiving advertising that is not relevant to the users. In order to achieve certain levels of personalization, online marketers make use of different third-party tracking tools. These third-party tracking tools may be considered by some as "privacy-invasive". 

In the following pages, we will present you with 7 short scenarios explaining how each category of third-party tracking tools work. Following each scenario, we will ask how you feel about the use of such tools being used on websites that you might commonly visit. We will also ask you whether you think it is okay for some websites to use such tools or not.

You must be at least 18 years of age to take part in this survey.
\subsection{Demographics}
What gender do you identify with?
\begin{enumerate}
    \item Female
    \item Male
    \item Other
    \item I prefer not to disclose this information
\end{enumerate}

What age group are you in?
\begin{enumerate}
    \item 18 - 30
    \item 31 - 45
    \item 46 - 60
    \item 60+
    \item I prefer not to disclose this information
\end{enumerate}

[We present the user with each scenario in the next section. The following section contains the three questions we ask about each of these scenarios.]

\subsection{Scenarios}
\label{app:scenarios}
\begin{enumerate}
\item A website X shares the fact that you are accessing website X, to company Y. Company Y adapts its advertisements on other websites based on the newfound fact that you have visited website X.
\item An adult website X shares the fact that you are accessing website X, to company Y. Company Y is an adult advertising company. Company Y pays other adult websites for space on their page. This lets company Y show you advertisements it thinks would be of interest to you. Company Y adapts its advertisements on other adult websites based on the newfound fact that you have accessed website X.
\item A website X uses company Y's service to identify your age, location, gender and device type when you are on their website X. Furthermore, website X is able to identify you when you access website X from a different device (e.g. smartphone or computer).
\item A website X uses company Y’s service to record all your mouse movements and everything you type into website X, so that website X can reconstruct how you interacted with the website (think of website X screen-recording your browsing session while on website X) in order to improve the usability of Website X.
\item A website X uses company Y's service to interact live with you through live-chat. When website X detects that you are idle (not moving your mouse) for some time, a chat-box appears on the right bottom corner of your screen, asking if you need assistance.
\item A website X includes a video player by video platform Y on their website X.  By loading website X's page, video platform Y knows that you are accessing website X. The next time you go on video platform Y, they suggest videos for you to watch based on the video you watched on website X.
\item A website X includes social media Y's "share" button to let you share their website X's content to your social media Y profile. However, even if you choose not to "share" anything, by loading website X's page, social media Y knows that you are accessing website X. The next time you go on social media Y, they show ads based on the newfound fact that you visited website X.
\item A website X uses company Y's service to provide a comment/review section on their website. The next time you go on another website using company Y's service, you could be prevented from posting comments if you posted an "undesirable" comment (e.g. spam) before on website X.
\end{enumerate}

\subsection{Questions for each scenario}
\label{app:questions}
\begin{enumerate}
    \item Before reading the above scenario, were you aware that companies do this type of tracking?
        \begin{itemize}
            \item Yes
            \item No
        \end{itemize}

    \item Based on the above scenario, how comfortable do you feel knowing that website X does this?
        \begin{enumerate}
            \item Very comfortable
            \item Somewhat comfortable
            \item Neutral
            \item Somewhat uncomfortable
            \item Very uncomfortable
        \end{enumerate}

    \item What types of websites do you think should not inform company Y of your visit, as described in the above scenario (check all that apply)?
        \begin{itemize}
            \item E-Commerce (online shop) [e.g. Under Armour]
            \item Social Media Platform [e.g. Facebook]
            \item News [e.g. CNN]
            \item Banking [e.g. Wells Fargo]
            \item Educational Institutions [e.g. MIT]
            \item Health Care [e.g. UPMC]
            \item Political [e.g. Donald J Trump]
            \item Non Governmental Organizations [e.g. PETA]
            \item Subscription-Based Services [e.g. Spotify]
            \item Adult Websites
            \item Other companies (non e-commerce) [e.g. Chili's]
            \item All of the above
            \item None of the above
    \end{itemize}

\end{enumerate}

You have reached the end of the survey. Thank you for your participation!

\section{Screenshots of the browser extension}
\label{app:sshots}

\begin{figure*}[!htbp]
\centering
\begin{subfigure}{.4\textwidth}
    \centering
  \includegraphics[width=60mm,height=120mm]{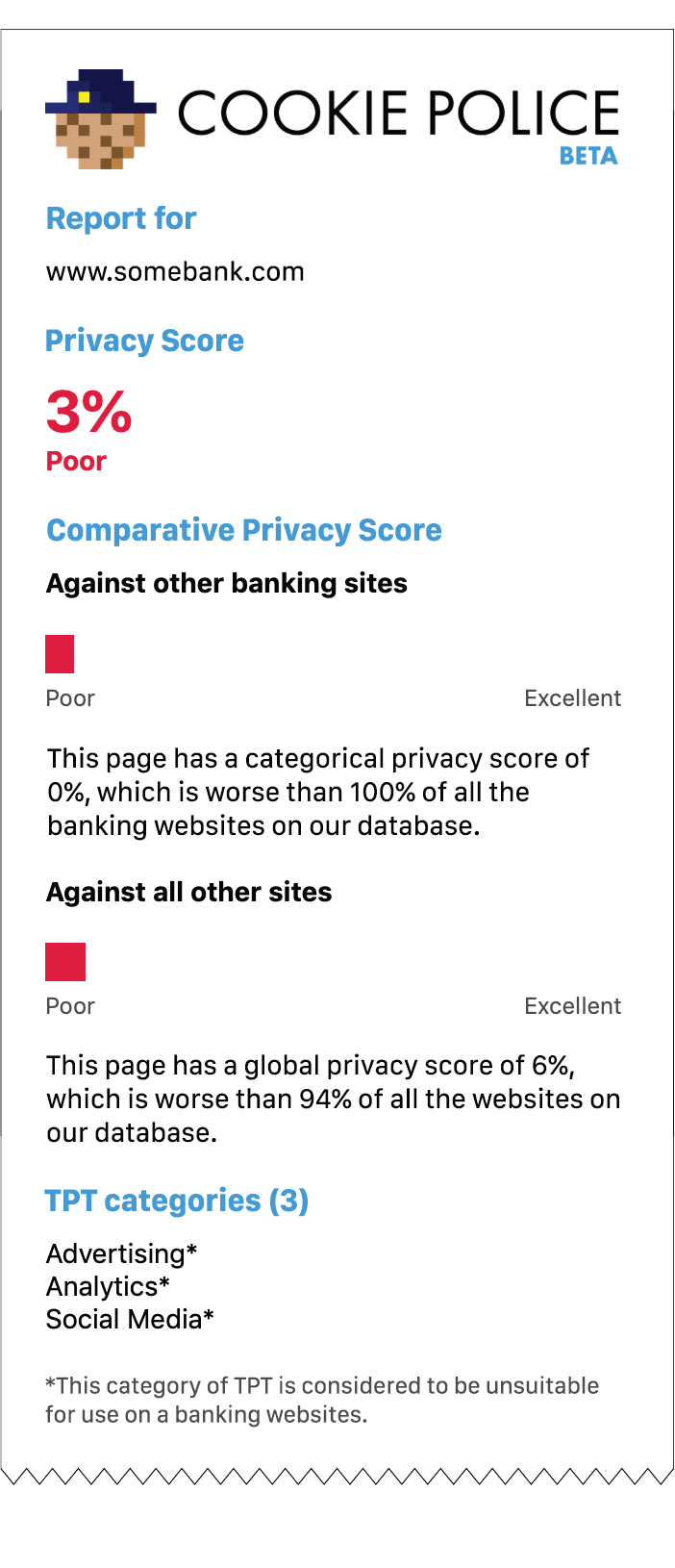}
  \label{fig:sub1}
\end{subfigure}
\begin{subfigure}{.4\textwidth}
  \centering
  \includegraphics[width=60mm,height=120mm]{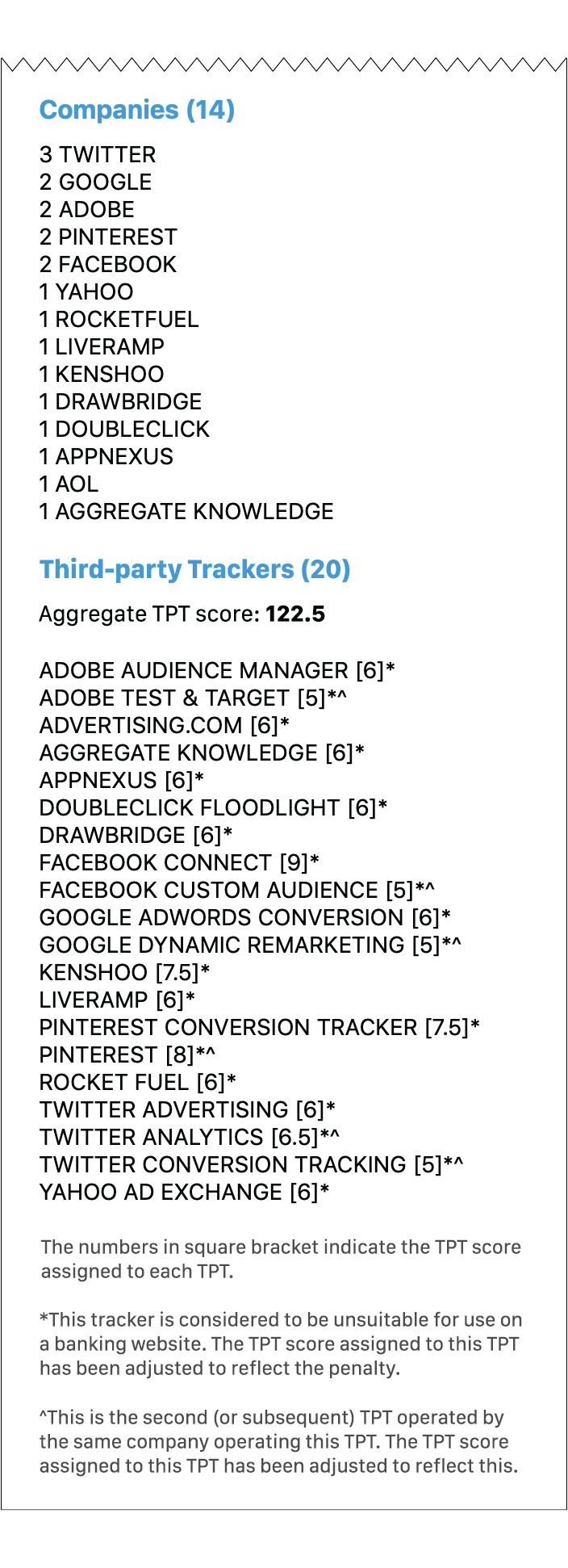}
  \label{fig:sub1}
\end{subfigure}
\caption{Screenshots of the extension's report \label{fig:sshots}}
\end{figure*}

Figure~\ref{fig:sshots} depicts screenshots of the
browser extension on a banking website.

%
% \centering
% \includegraphics[width=0.4\textwidth]{figures/updatedss1.png}
% \caption{Screenshot of extension prototype \label{fig:cp_ss1}}
% \end{figure*}

% \begin{figure*}[]
% \centering
% \includegraphics[width=0.4\textwidth]{figures/updatedss2.png}
% \caption{Screenshot of extension prototype (continued) \label{fig:cp_ss2}}
% \end{figure*}

\end{document}